\documentclass[preprint]{ptephy_v1}

\preprintnumber{arXiv:1807.06753} 
\usepackage{hyperref}
\usepackage{amsmath,amssymb,array}
\usepackage{bm}
\usepackage{delarray}

\newcommand{\ket}[1]{| { #1} \rangle}
\newcommand{\bra}[1]{ \langle {#1} |}

\def \a#1{{${\hat a}_#1$}}

\def \a#1{\textcolor{black}{#1}}

\begin{document}

\title{Do black holes store negative entropy?}


\author{Koji Azuma}
\affil{NTT Basic Research Laboratories \& NTT Research Center for Theoretical Quantum Information, NTT Corporation, 3-1 Morinosato Wakamiya, Atsugi, Kanagawa 243-0198, Japan \email{koji.azuma@ntt.com}}

\author{Sathyawageeswar Subramanian}
\affil{
Department of Computer Science and Technology,
University of Cambridge,
15 JJ Thomson Avenue
Cambridge CB3 0FD
}

\author{Go Kato}
\affil{Advanced ICT Research Institute, NICT
4–2–1, Nukui-Kitamachi, Koganei, Tokyo 184-8795, Japan}


\begin{abstract}
The Bekenstein-Hawking equation states that black holes should have entropy proportional to their areas to make black hole physics compatible with the second law of thermodynamics. However, this equation leads to an inconsistency among the first law of black hole mechanics, the entropy conservation law of quantum mechanics, and a heuristic picture for Hawking radiation---creation of entangled pairs near the horizon.
Here we propose an equation alternative to the Bekenstein-Hawking equation from the viewpoint of quantum information, to resolve this inconsistency without changing Hawking's original pair-creation picture for the radiation.
This argues that the area of any stationary black hole, including Kerr and charged ones, is proportional to the coherent information---which is `minus' the conditional entropy defined only in the quantum regime---from the outside, to the black hole excluding negative-frequency particles generated by Hawking's pair creation. Our equation suggests that negative-frequency particles inside a black hole behave as if they have `negative' entropy. Our result implies that a black hole stores purely quantum information, rather than classical information, and the area of the event horizon describes the number of Bell pairs that can be distilled between the interior and exterior.
\end{abstract}

\subjectindex{A61,B30,E10}

\maketitle

\section{Introduction}
Black holes are one of the most beautiful but mysterious objects in our universe. Although their original conception came about as purely theoretical objects encountered as solutions of the Einstein equation in general relativity, nowadays, it is a target of observational astrophysics \cite{E19-1,E19-2,E19-3,E19-4,E19-5,E19-6,A16}.
For several years, the typical
picture that black holes can only absorb and nothing can escape from them, looked highly irreversible, compared with normal stars. However, this is merely a viewpoint for black holes in the regime of classical general relativity, and turns out not to be the case for the quantum world. In particular, remarkably, Hawking has developed a semi-classical picture \cite{H74,H75} where thermal radiation occurs from a Schwarzschild black hole---although notably, Schwarzschild black holes are regarded as `useless' classically because we cannot distil or extract energy from them, in contrast to Kerr or charged black holes \cite{P69,MTW}. As a result, Hawking famously described it by saying `{\it a ``black hole'' is not completely black}' \cite{H76}. However, this Hawking radiation gave us many rich and serious puzzles to consider, about the consistency between such black hole mechanics and quantum mechanics.

A puzzle appears \cite{BPZ13,M09,M19} when we combine the ideas of Hawking radiation with the first law of black hole mechanics \cite{BCH73,MTW}, the Bekenstein-Hawking equation \cite{B73,B74,H74,H75,H76} and quantum mechanics. The first law of black hole mechanics is associated with conservation laws: for a stationary black hole $B$, we have
\begin{equation}
\label{eq:firstlaw}
{\rm d}(M_B c^2)=\frac{\kappa_Bc^2}{8\pi G}{\rm d}A_B + \Omega_B {\rm d}J_B + \phi_B {\rm d}Q_B, 
\end{equation}
where $c$ is the speed of light, $G$ is the Newtonian constant of gravitation,
$M_B$ is the rest mass, $A_B$ is the area of the event horizon, $\kappa_B$ is the surface gravity, $J_B$ is the angular momentum, $\Omega_B$ is the angular velocity, $Q_B$ is the charge and $\phi_B$ is the electrostatic potential of the black hole (see, e.g., Ref. \cite{H76}, for their explicit forms). 
Here $\Omega_B {\rm d}J_B + \phi_B {\rm d}Q_B$ in the first law corresponds to the change of black hole energy in the form of work done, and thus,
\begin{equation}
\delta {\cal Q}_B := {\rm d}(M_B c^2)- \Omega_B {\rm d}J_B - \phi_B {\rm d}Q_B \label{eq:heat}
\end{equation}
can be deemed to be the change of `heat', which does not decrease classically  \cite{P69,MTW}.
On the other hand,
the Bekenstein-Hawking equation is an area law for the black hole:
\begin{equation}
{\rm d} S(B)=\frac{c^3}{4G \hbar} {\rm d}A_B, \label{eq:Beken}
\end{equation}
where $\hbar:=h/(2 \pi)$ with the Planck constant $h$, and $S(B)$ is (dimensionless) entropy of the black hole $B$. 
Even if we assume that the entropy $S$ represents the von Neumann entropy which is defined by $S(\hat{\rho}):=-{\rm Tr}[\hat{\rho}\ln \hat{\rho}]$ for density operators $\hat{\rho}$, this law itself can be compatible with Hawking's area theorem \cite{H72} in general relativity and Bekenstein's generalized second law \cite{B74}, even in the quantum mechanical point of view (see Appendix~\ref{sec:A1}).
In contrast, those laws (\ref{eq:firstlaw}) and (\ref{eq:Beken}) are inconsistent with the pair-creation picture of Hawking radiation, from the quantum mechanical point of view \cite{M09,BPZ13,M19}. More precisely, Hawking's finding is that an observer at the future infinity receives mass-less scalar bosons $H^+$ in a thermal state with Hawking temperature $T_H:=\kappa_B \hbar/(2 \pi k c)$ (where $k$ is the Boltzmann constant) from a Schwarzschild black hole (with $\Omega_B=0$ and $\phi_B=0$). Its purification partner $H^-$---which must exist somewhere in our universe in the quantum mechanical description used by the observer at future infinity, according to the concept of purification \cite{NC}---is regarded as bosons having a {\it negative} frequency and thus appearing in the interior of the black hole $B$ through quantum tunnelling (as will be explained in Sec.~\ref{se:entropy-cons}). The fact that the observer receives the positive energy of the radiation $H^+$ means ${\rm d} M_B<0$ from the energy-conservation law, implying ${\rm d}A_B <0$ according to the first law (\ref{eq:firstlaw}) with $\Omega_B=0$ and $\phi_B=0$. On the other hand, quantum mechanics for the observer at the future infinity tells us that the purification partner $H^-$ has the same {\it positive} entropy as the thermal radiation. That is, $S(H^-)=S(H^+)>0$ with any unitary-invariant entropic measure $S$ (e.g., the von Neumann entropy) because $H^+H^-$ is in a pure state. Thus, the fact that the black hole receives this purification partner $H^-$ with {\it positive} entropy means ${\rm d}S(B) >0$, implying ${\rm d}A_B >0$ if the entropy $S(B)$ in the Bekenstein-Hawking equation (\ref{eq:Beken}) simply represents such unitary-invariant entropy, say the von Neumann entropy. Hence, Hawking's original pair-creation picture for Hawking radiation leads to a paradox about the direction of the change of the area $A_B$.

Perhaps, one might think that if $S(B)$ in Eq.~(\ref{eq:Beken}) is regarded as the von Neumann entropy of a system $B$ in a maximum-entropy ensemble, i.e., the coarse-grained entropy, the above paradox does not appear, because fundamental thermodynamic relation, such as ${\rm d} (M_Bc^2)= T_H {\rm d}S(B)$ in the case of Schwarzschild black holes, follows in this case. 
However, it might be challenging to develop a complete microscopic theory along this line.
For example, if $S(B)$ in Eq.~(\ref{eq:Beken}) is associated with the coarse-grained entropy \cite{AHMST20} of a system $B$ with a constraint that the expectation value of a Hamiltonian is $M_Bc^2$, that is, the von Neumann entropy of a canonical ensemble of the system $B$,
the heat capacity must be positive, because the heat capacity of any canonical ensemble is directly proportional to the variance of the Hamiltonian (which is always positive). 
However, the heat capacities of normal black holes are negative [in contrast to black holes in anti-de Sitter (AdS) spaces whose heat capacities are typically positive]. 
Hawking also pointed out the difficulty of associating the state of a black hole with a canonical ensemble, stemming from this negativity of the heat capacity \cite{H76}. One might think that this problem may not occur if we regard the state of the black hole as a microcanonical ensemble. In fact, this is normally assumed \cite{H76,H90,SV96,SL04}, referred to as a `central dogma' \cite{AHMST20}. However, notice that this is merely a postulate. Besides, 
there is a serious problem for the use of statistical mechanics itself, if we believe Hawking's pair-creation picture is correct. In fact, since, in the pair-creation picture, the black hole system $B$ can store bosons  $H^-$ with a negative frequency, any coarse-grained entropy, including one based on a microcanonical ensemble, cannot be defined, because the system should have not only normal objects, such as ones composing its positive mass and positive frequencies,  
but also negative-frequency bosons $H^-$ with an unbounded negative energy spectrum. 

A way to resolve this paradox is to regard Hawking's pair-creation picture as heuristic at best---accompanied with the existence of other confusions (see, e.g.,  \cite{H16})---and to explore other mechanisms for the radiation from a black hole (see, e.g.,  \cite{PW00,H16,BPZ13,BP11}). For example, in \cite{PW00, BPZ13,BP11}. the radiation from a hole is associated with tunnelling a particle {\it inside} the hole to the outside, to decrease the dimensionality (and thus, the von Neumann entropy) of the hole through the radiation. 
However, this type of other mechanism tends to fling away a good selling point, as well, of Hawking's pair-creation picture that directly associates the radiation temperature with the surface gravity $\kappa_B$ of the hole.

In this paper, we present an alternative idea to resolve the above paradox without changing Hawking's original pair-creation picture, just by assuming that any stationary black hole with Hawking radiation stores quantum entanglement, rather than simple entropy such as the von Neumann entropy or the coarse-grained entropy based on statistical mechanics. More precisely, we postulate that 
\begin{equation}
\frac{c^3}{4G \hbar}  {\rm d}A_B = {\rm d}I(\bar{B} \rangle B^+)={\rm d}S(B^+)-{\rm d}S(B^-)\label{eq:m1}
\end{equation}
holds for any quasi-static process of the black hole $B$,
where $I(X \rangle Y):=-S(X|Y)$ is called the coherent information \cite{SN96,HOW05,HOW07} from $X$ to $Y$ and $S (X|Y)$ is the conditional entropy defined by $S(X|Y):=S(XY)-S(Y)$ with the von Neumann entropy $S(X):=S(\hat{\rho}_X):=-{\rm Tr}[\hat{\rho}_X\ln \hat{\rho}_X]$ for a system $X$ in a state $\hat{\rho}_X$. 
The coherent information can be positive only in the quantum world and it is associated with entanglement; it has a clear information-theoretic meaning in the context of entanglement distillation \cite{DW05} (see Ref.~\cite{HHHH09} for its context).
In the modified area law~(\ref{eq:m1}),
we assume the following:
(i) the black hole $B$ includes `negative' subsystem $B^-$ that is constituted only of negative-frequency particles $H^-$ generated by Hawking's pair creation, and its relative complement with respect to the black hole $B$ is referred to as `positive' subsystem $B^+$ in this paper (in the sense that ingredients of this part $B^+$ could change even in classical general relativity without considering Hawking radiation), i.e. $B=B^+B^-$ (more precisely, ${\cal H}_B = {\cal H}_{B^+} \otimes {\cal H}_{B^-}$, where ${\cal H}_X$ is the Hilbert space for a system $X$); (ii) the whole system $B\bar{B}=B^+B^-\bar{B}$, by including a system $\bar{B}$ in the outside of the black hole $B$, can always be in a pure state (although ${\cal H}_{B} \otimes {\cal H}_{\bar{B}}={\cal H}_{B^+}\otimes {\cal H}_{B^-} \otimes {\cal H}_{\bar{B}}$ is a Hilbert subspace of the universe in general), which leads to the final equality in Eq.~(\ref{eq:m1}) from $S(\bar{B}B^+)=S(B^-)$; (iii) the free evolution of the black hole $B$ is described by a unitary operator in the form $\hat{U}_{B^+} \otimes \hat{V}_{B^-}$.

The assumption (i) should hold for a static observer at spatial infinity outside the black hole who 
infers the appearance of purification partner $H^-$ inside it indirectly through the observation of Hawking radiation $H^+$. 
The assumption (ii) states that the purification partner $\bar{B}$ of the state of the black hole $B$ exists outside its event horizon, according to quantum mechanics. 
The assumption (iii) is a sufficient condition for the area $A_B$ of the black hole $B$ to be unchanging in its free evolution, under Eq.~(\ref{eq:m1}). Indeed, the area $A_B$ in Eq.~(\ref{eq:m1}) would be unaltered even for unitarily pair creation or annihilation of an entangled {\it pure} state between $B^+$ and $B^-$ (because ${\rm d}S(B^+)={\rm d}S(B^-)$ holds in this case), although such a process would be represented by a global unitary operation beyond local unitary operations of the form $\hat{U}_{B^+} \otimes \hat{V}_{B^-}$. Nevertheless, we stick to the assumption (iii)\footnote{
The assumption (iii) is also supported by a pair-creation picture \cite{H96} summarised by 't Hooft for a Schwarzschild black hole, where negative-frequency particles $H^-$ of Hawking pairs emerge as ordinary particles in the second asymptotically flat region of
the maximally extended Schwarzschild spacetime, rather than in the region usually identified as the black hole's interior (having the singularity). This view is used in 't Hooft's quantum unitary description of the black-hole physics \cite{H16a,H16b}. In this context, $B^-$ and $B^+$ could be identified as the second asymptotically flat region and the black hole's interior region, respectively, supporting the assumption (iii) from the existence of an event horizon between them. } because it is more reasonable than blindly assuming a form of a unitary interaction inside the hole between infallen normal particles and unconventional tunnelled negative-frequency {\it bosons} (to which we cannot apply a notion like `Dirac's negative energy sea', implying that we are not allowed to replace them with corresponding antiparticles).

Notice that our equation (\ref{eq:m1}) coincides with the Bekenstein-Hawking equation (\ref{eq:Beken}) as long as the effect of the Hawking radiation is very small (i.e., ${\rm d}S(B^-) \ll {\rm d}S(B^+)$), which reproduces all known results shown with the original equation (\ref{eq:Beken}), such as Bekenstein's generalised second law \cite{B74}.
Nevertheless, Eq.~(\ref{eq:m1}) not only solves the paradox above in contrast to Eq.~(\ref{eq:Beken}), but also sheds new light on other paradoxes such as the information loss paradox \cite{H76b,P92} and the firewall paradox \cite{M09,AMPS13}.

\section{Entropy conservation and trajectory}\label{se:entropy-cons}
Let us start by explaining how to evaluate the right-hand side of Eq.~(\ref{eq:m1}) for a black hole $B=B^+B^-$ composed of the positive subsystem $B^+$ and the negative subsystem $B^-$. This can be done by using the following rule from the entropy conservation law in quantum mechanics:
if a stationary black hole $B=B^+B^-$ stores an infalling positive (or tunnelling \cite{H75} negative) object $C$ which is defined as one having positive (negative) `kinetic' energy $K_C:=E_C-\Omega_B l_C-\phi_B Q_C\ge0$ ($K_C < 0$) with energy-at-infinity $E_C$ \cite{MTW}, axial component $l_C$ of angular momentum and charge $Q_C$---evaluated at event of crossing, then the entropy changes $\Delta S(B^\pm):=S(B'^\pm)-S(B^\pm)$ of the positive and negative parts of the black hole are $\Delta S(B^+)=S(C)$ and $\Delta S(B^-)=0$ ($\Delta S(B^-)=S(C)$ and $\Delta S(B^+)=0$).
Here, the purification partner of the object $C$ is assumed to exist in the outside $\bar{B}$, 
$S(C)$ is the von Neumann entropy of the system $C$ at that event, and $B'=B'^+ B'^-$ is the black hole after capturing the object $C$ in its positive (negative) part at the vicinity of the horizon and then evolving unitarily in each part according to a unitary operator $\hat{U}_{B'^+} \otimes \hat{V}_{B'^-}$, i.e., ${\cal H}_{B'^+}={\cal H}_{B^+} \otimes {\cal H}_C$ and ${\cal H}_{B'^-}= {\cal H}_{B^-}$ (${\cal H}_{B'^-}={\cal H}_{B^-} \otimes {\cal H}_C$ and ${\cal H}_{B'^+}= {\cal H}_{B^+}$). For each case, $\Delta S(B^\pm)=S(C)$ and $\Delta S(B^\mp)=0$ follow from $S(B'^\pm) =S(B^\pm C)=S(B^\pm)+S(C) $ and $S(B'^\mp)=S(B^\mp)$, where we have used the invariance of the von Neumann entropy under any unitary operation (i.e., $S(B'^\pm) =S(B^\pm C)$ and $S(B'^\mp)=S(B^\mp)$ under the unitary $\hat{U}_{B'^+} \otimes \hat{V}_{B'^-}$) and $I(B^\pm:C)=0$ for the mutual information $I(X:Y):=S(X)+S(Y)-S(XY)$ from the condition for the purification partner of the object $C$.

In the rule, it is assumed that object $C$ is very small, that is, its size and mass are much smaller than those of the hole and it has sufficiently small charge, so that its gravitational/electromagnetic radiation is negligible. Thus, it moves very nearly along a test-particle trajectory, which approaches the horizon with future-pointing 4-momentum when $K_C \ge 0$, or with past-pointing 4-momentum when $ K_C \le 0$ (corresponding to `negative root states') (see Sec.~33.7 of Ref.~\cite{MTW}). 
The negative object $C$ (with $ K_C \le 0$) in the rule can be composed only of the purification partner $H^-$ of Hawking radiation $H^+$, tunnelling into the inside of the hole $B$ as a quantum effect,
because there is no test particle $C$ in the realm of classical general relativity that crosses the horizon with $K_C<0$ as noted in Sec.~33.7 of Ref.~\cite{MTW}. On the other hand, the positive object $C$ (with $K_C \ge 0$) in the rule represents a test particle generated in the outside $\bar{B}$ whose orbit can cross the horizon, including \a{normal} particles treated in the framework of classical general relativity (because of $K_C\ge 0$). Therefore, the \a{negative} object $C$ can belong to the \a{negative} subsystem $B^-$ only, while the positive object $C$ can interact with its relative complement $B^+$ only.
The energy-at-infinity
$E_C$, angular momentum $l_C$, charge $Q_C$ and von Neumann entropy $S(C)$ in the above rule should be determined once we are given a density operator $\hat{\rho}_C$ which is the quantum description of the internal state of system $C$ at event of crossing. This is similar to the treatment studied in Ref.~\cite{B73}.
Therefore, in this framework, the rule for entropy is analogous to the energy conservation law in general relativity (e.g., see Sec.~33.7 of Ref.~\cite{MTW}).

\section{Equivalence between the modified area law (\ref{eq:m1}) and the first law (\ref{eq:firstlaw})}
Here we show our main result that the modified area law (\ref{eq:m1}) is equivalent to the first law (\ref{eq:firstlaw}) for a stationary black hole, in contrast to the Bekenstein-Hawking area law (\ref{eq:Beken}), even if a black hole emits Hawking radiation continuously, following Hawking's original pair-creation picture.
Let us consider a stationary black hole $B$, which emits a Hawking pair $H^+H^-$ in state 
\begin{align}
\ket{\chi}_{H^+ H^-} := \exp[r_{\omega'}(\hat{a}_k^\dag \hat{b}_{-k}^\dag- \hat{a}_k\hat{b}_{-k})]\ket{{\rm vac}} 
=  \frac{1}{\cosh r_{\omega'}} \sum_{n=0}^{\infty} \tanh^n r_{\omega'} \ket{n}_{H^+} \ket{n}_{H^-} ,\label{eq:two-mode}
\end{align}
where $\hat{a}_k$ and $\hat{b}_{-k}$ are annihilation operators for massless scalar bosons associated with the positive-frequency particles $H^+$ and the negative-frequency particles $H^-$ respectively, the parameter $r_{\omega'}$ is related \cite{H76} to a mode with frequency $\omega(>0)$, angular momentum $m\hbar$ about the axis of rotation of the black hole, and charge $e$ via 
$ \omega':=  \omega- m  \Omega_B -e \phi_B/\hbar(>0)$
and
$\exp(-\pi c\omega' /\kappa_B)=\tanh r_{\omega'}$, 
and the effective mode frequency $\omega'$ will follow some dispersion relation $\omega'=\omega'(\pm k)$. 
In the pair creation picture, the negative-frequency particles ${H}^-$ appear in a mode tunnelling into the black hole (i.e. on a worldline crossing the event horizon), while the positive-frequency particles ${ H}^+$ appear in a mode propagating from the vicinity of the event horizon to a static observer at infinity. 
The reduced density operator of the positive-frequency particles ${ H}^+$ is the Gibbs state with the inverse temperature~$\beta_{ H}:=(k T_{H})^{-1}$,
\begin{align}
\hat{\chi}_{{ H}^+}:=& {\rm Tr}_{H^-} [\ket{\chi}\bra{\chi}_{H^+H^-}]   
=\frac{1}{\cosh^2 r_{\omega'}} \sum_{n=0}^\infty \tanh^{2n}r_{\omega'} \ket{n}\bra{n}_{{H}^+} 
= \frac{e^{-\beta_{H} \hbar \omega' \hat{n}_{{H}^+} }}{Z_{\beta_{H} }}, \label{eq:chi}
\end{align}
where $\hat{n}_{{ H}^+}:=\hat{a}_k^\dag \hat{a}_k$ and $Z_{\beta_{ H}}:=(1-e^{-\frac{2\pi c  \omega' }{\kappa_B}})^{-1}=(1-e^{-\beta_{ H} \hbar \omega'})^{-1}$ is the partition function.
Since this satisfies
$
- \ln \hat{\chi}_{{ H}^+}= \beta_{ H} \hbar \omega' \hat{n}_{{ H}^+} +\ln Z_{\beta_{ H}} \hat{1}_{{ H}^+}
$, we have
\begin{equation}
\label{eq:Gibbs_S}
S({ H}^+)=\beta_{ H}  \hbar \omega' n_{{ H}^+} + \ln Z_{\beta_{ H}} =\beta_{ H}  K_{H^+} + \ln Z_{\beta_{ H}} ,
\end{equation}
where
\begin{align}
n_{{ H}^+} &={\rm Tr}[ \hat{n}_{ H^+}  \hat{\chi}_{{ H}^+} ] = \frac{1}{e^{\beta_{ H}\hbar \omega'}-1} ,\label{eq:nH+}\\
K_{H^+} &=\hbar \omega' n_{{ H}^+} = \hbar \omega n_{{ H}^+} - m \hbar \Omega_B n_{{ H}^+} - e\phi_B n_{{ H}^+}(\ge 0). \label{eq:KH+}
\end{align}
Hence, the positive-frequency particles $H^+$ satisfy
\begin{multline}
\frac{{\rm d} S( H^+) }{  {\rm d} K_{H^+}}=
\frac{1}{\hbar\omega'} \frac{{\rm d} S( H^+) }{  {\rm d} n_{{ H}^+}} 
=\frac{1}{\hbar \omega'} \biggl( \beta_{ H} \hbar \omega' + \hbar \omega' n_{{ H}^+  } \frac{{\rm d} \beta_{ H}} {{\rm d} n_{{ H}^+}}  
 + \frac{1}{Z_{\beta_{ H}}} \frac{\partial Z_{\beta_{ H}}}{\partial \beta_{ H}} \frac{{\rm d} \beta_{ H}} {{\rm d} n_{{ H}^+}} \biggr) 
=\beta_{ H} \label{eq:deriv}
\end{multline}
for given $\omega'$. Therefore, the emission of positive-frequency particles $H^+$ from the event horizon is pure thermal radiation at the Hawking temperature $T_H$. 
Note that Eq.~(\ref{eq:deriv}) implies that ${\rm d}K_{H^+}$ is the received heat $\delta {\cal Q}_{H^+}$ of positive-energy particles $H^+$ via a quasi-static process for the system $H^+$ (i.e., ${\rm d}K_{H^+}=\delta {\cal Q}_{H^+}$), because $H^+$ is initially in a thermal equilibrium state and $\beta_{ H}^{-1}{\rm d} S( H^+) =\delta {\cal Q}_{H^+}$ holds for any quasi-static process. Besides, notice that $H^-$ with the number operator $\hat{n}_{{ H}^-}:= \hat{b}_{-k}^\dag \hat{b}_{-k}$ are negative-frequency particles with $K_{H^-}=-\hbar \omega' n_{H^-}=-\hbar \omega' n_{H^+}=-K_{H^+}\le0$ \cite{H96}, which are associated with a `negative root state' in Sec.~33.7 of Ref.~\cite{MTW}.

Now, we consider a process where the black hole emits the Hawking radiation $H^+$ to infinity, while, from the outside, it absorbs an infalling positive object $C$ with positive kinetic energy $K_C(\ge 0)$ and entropy $S(C)$.
Here the infalling system $C$ is initially decoupled with the black hole $B$ before this absorption, implying that its purification partner belongs to the outside $\bar{B}$.
In this process, the black hole $B$ loses energy $K_{H^+}$ of positive-frequency particles $H^+$ as heat but receives entropy $S(H^-)$ of negative-frequency particles $H^-$ through the Hawking radiation, while it receives kinetic energy $K_C$ as heat and entropy $S(C)$ by absorbing the positive object $C$. Therefore, in this process, hole's heat change $\Delta {\cal Q}_B$ of Eq.~(\ref{eq:heat}) and the change $\Delta I(\bar{B}\rangle B^+)$ of coherent information are given by
\begin{align}
&\Delta {\cal Q}_B = K_C -K_{H^+}, \label{eq:ch-e} \\
&\Delta I(\bar{B}\rangle B^+) = S(C) -  S(H^-) = S(C) -  S(H^+), \label{eq:ch-I}
\end{align}
where we have used conservation laws for energy, charge, and axial component of angular momentum (see Sec.~33.7 of Ref.~\cite{MTW}) in Eq.~(\ref{eq:ch-e}), as well as $S(H^-)=S(H^+)$ for the pure state $\ket{\chi}_{H^+H^-}$ in Eq.~(\ref{eq:ch-I}). At this point, notice that although $\Delta I(\bar{B}\rangle B^+)$ can take both positive and negative values similarly to $\Delta {\cal Q}_B$ (which could take even negative values due to Hawking radiation), $\Delta S(B)= S(C)+S(H^-) \ge S(H^-)  > 0$ always holds in this process\footnote{This increase of $S(B)$ that can occur even when $S(C)=0$ is also pointed out by Brauntein {\it et al.} \cite{BPZ13} as
`{\it pair creation necessarily requires
the dimensionality of the interior Hilbert space of a black
hole to be increasing while simultaneously its physical size
is decreasing}'.
}, implying $\Delta A_B > 0$ according to the Bekenstein-Hawking equation (\ref{eq:Beken}).
This is one underlying reason why the Bekenstein-Hawking equation (\ref{eq:Beken}) cannot explain the existence of a stationary black hole with $\Delta A_B=0$ under the occurrence of Hawking radiation, let alone the area decrease $\Delta A_B < 0$ caused by the Hawking radiation according to the first law (\ref{eq:firstlaw}).

Since the first law (\ref{eq:firstlaw}) holds for stationary black holes, to refer to the law, we first need to make a setup in which a black hole is stationary even under the existence of Hawking radiation, irrespectively of the sign of the heat capacity. A way to do so is to assume that the positive object $C$ is a bosonic system with the effective mode frequency $\omega'$ (which is the same as one of the system $H^+$) and is in the same state as the radiation $H^+$, that is, in a thermal state $\hat{\chi}_C$ as in Eq.~(\ref{eq:chi}), with the Hawking temperature $T_{H}$. 
Then, we have 
\begin{equation}
\hbar \omega' n_C=\hbar \omega' n_{H^+}, \label{eq:eq}
\end{equation}
meaning $K_C=K_{H^+}$ from Eq.~(\ref{eq:KH+})
and $S(H^+)=S(C)$ from Eq.~(\ref{eq:Gibbs_S}).  Hence, in this case, the above process provides $\Delta {\cal Q}_B=0$ and $\Delta I(\bar{B}\rangle B^+)=0$ from Eqs.~(\ref{eq:ch-e}) and (\ref{eq:ch-I}), from which we conclude $\Delta A_B=0$, using either from the first law (\ref{eq:firstlaw}) for stationary black holes or the modified area law (\ref{eq:m1}).
Therefore, as long as this equilibrium process is repeated, say if a black hole repeatedly interacts with thermal systems with the Hawking temperature $T_H$, the black hole can be exactly in a stationary state.

Let us move on to a case where the above equilibrium process is repeated, but at some point, it deviates slightly from its equilibrium version, accompanied by small changes on the system $C$ and the Hawking radiation $H^+$ such that $K_C: \hbar \omega' n_C \to \hbar\omega' n_C  +\hbar \omega' \Delta n_C$ and $K_{H^+} : \hbar\omega' n_{H^+} \to \hbar\omega' n_{H^+}  +\hbar\omega' \Delta n_{H^+}$. 
For this perturbation, from Eqs.~(\ref{eq:KH+}) and (\ref{eq:eq}), Eq.~(\ref{eq:ch-e}) becomes
\begin{align}
\Delta {\cal Q}_B =& \hbar\omega' n_C  + \hbar \omega' \Delta n_C -( \hbar\omega' n_{H^+}  +\hbar\omega' \Delta n_{H^+}) \nonumber \\
=&\hbar \omega'  \Delta n_C  -\hbar \omega' \Delta n_{H^+}, \label{eq:dE}
\end{align}
while Eq.~(\ref{eq:ch-I}) becomes
\begin{align}
\Delta I(\bar{B}\rangle B^+) =& \beta_{ H}  \hbar \omega' n_{C} + \ln Z_{\beta_{ H}}+\Delta S(C) 
- (\beta_{ H} \hbar \omega' n_{{ H}^+} + \ln Z_{\beta_{ H}}+\Delta S(H^+)) \nonumber \\
=& \Delta S(C) -\Delta S(H^+),  \label{eq:dI}
\end{align}
using Eq.~(\ref{eq:Gibbs_S}).
As long as the perturbation is small enough to be regarded as a quasi-static process for system $C$ and Hawking radiation $H^+$, the difference $\Delta X$ on a quantity $X$ can be regarded as its derivative and Eq.~(\ref{eq:deriv}) should hold. Hence, we have 
\begin{align}
{\rm d}I(\bar{B}\rangle B^+) =& {\rm d} S(C) -{\rm d} S(H^+) 
= \beta_H (\hbar \omega' {\rm d} n_C -\hbar \omega' {\rm d} n_{H^+}) \nonumber \\
=& \beta_H \delta {\cal Q}_B   
\label{eq:deri}
\end{align}
from Eqs.~(\ref{eq:dE}) and (\ref{eq:dI}).
Combined with the first law (\ref{eq:firstlaw}) for stationary black holes, this concludes Eq.~(\ref{eq:m1}) (although $\delta {\cal Q}_B$ can now be either positive or negative, depending on fluctuations of systems $H^+$ and $C$, in contrast to purely classical scenarios).
Notice that this proof works irrespectively of whether the heat capacity of the black hole is positive or negative. This is in contrast to methods used for AdS black holes \cite{AHMST20,H16}.

\section{Implications of the modified area law (\ref{eq:m1})}

Since we have shown the equivalence between the modified area law (\ref{eq:m1}) and the first law (\ref{eq:firstlaw}),
here we elucidate several important properties of the modified area law (\ref{eq:m1}). As we have already noted, our expression (\ref{eq:m1}) for the area of a black hole is invariant under any unitary operation of the form $\hat{U}_{B^+}\otimes \hat{V}_{B^-} \otimes \hat{W}_{\bar{B}}$.
This means that the area of the black hole is unaltered unless interactions occur among positive part $B^+$ and negative part $B^-$ of the black hole $B$ and its outside $\bar{B}$.
However, remember that this is merely a sufficient condition for the area $A_B$ to be unchanging: more generally, any process with ${\rm d}S(B^+)={\rm d}S(B^-)$ does not change its area, according to Eq.~(\ref{eq:m1}).

Another property is the reduction of our expression (\ref{eq:m1}) to the Bekenstein-Hawking equation (\ref{eq:Beken}) for any black hole dynamics unaccompanied by any change of the negative-frequency particles, which leads to Bekenstein's generalised second law \cite{B74}:
\begin{equation}
\frac{{\rm d} A_B}{4} +{\rm d}S(\bar{B})\ge0. \label{eq:generalized2}
\end{equation}
This is because such dynamics satisfy ${\rm d} S(B^-)=0$, and Bekenstein's consideration is based on the assumption that black holes are composed only of normal objects (that is, entropy $S(B)$ in Eq.~(\ref{eq:Beken}) is equivalent to $S(B^+)$ in our expression (\ref{eq:m1})). 

The only dynamics which cannot be explained by this reduction are the increase of the area through black hole mergers from Hawking's area theorem \cite{H72}, because we have modelled black holes as being composed of not only normal objects, but also negative-frequency ones. For this case, we regard the merger between two black holes $B_1$ and $B_2$ as occurring via some isometric dynamics $\hat{U}_{B_1^+ B_2^+ \to B^+G}\otimes \hat{V}_{B_1^- B_2^-\to B^-}$, to form a new black hole $B$. For simplicity of exposition, the initial black holes are assumed to be decoupled, i.e., $I(B_1^+:B_2^+)=0$ and $I(B_1^-:B_2^-)=0$, and $G$ represents a system emitted into the universe (the outside $\bar{B}$) as back reaction (e.g. gravitational waves). Then, from Eq.~(\ref{eq:m1}), we have
\begin{align}
I(\bar{B}\rangle B^+)=&S(B^+)-S(B^-) 
= S(B^+ G) -S(B^-) -S(G|B^+)  \nonumber \\
=& S(B_1^+B_2^+) - S(B_1^-B_2^-) -S(G|B^+) \nonumber \\
=& I(\bar{B}_1\rangle B_1^+)+I(\bar{B}_2\rangle B_2^+) -S(G|B^+),
\end{align}
where $\bar{X}$ is the complement of system $X$, that is, all the related systems except for system $X$.
Therefore, if the system $G$ is in either a pure state or an entangled state so as to have negative conditional entropy ($S(G|B^+) \le 0$), the net black hole area increases, in accordance with our equation (\ref{eq:m1}). In particular, if initial black holes $B_1$ and $B_2$ and the final black hole $B'$ are stationary and the integration constant of Eq.~(\ref{eq:m1}) is zero, we have $A_{B}\ge A_{B_1} +A_{B_2}$. Since gravitational waves can tell us of the existence of black hole mergers \cite{A16}, it may not be unnatural for the emitted system $G$ to be highly entangled with the positive part $B^+$ of the black hole $B$. Besides, note that the condition $S(G|B^+) \le 0$ is essentially the same as the one necessary for the black hole merger to satisfy Hawking's area theorem \cite{H72}, even with the Bekenstein-Hawking equation (\ref{eq:Beken}) (see Appendix~\ref{sec:A1}).

Finally, we point out an operational meaning of the area $A_B$ of our equation (\ref{eq:m1}), from a point of view that arises exclusively in quantum information theory.
In quantum information theory, the coherent information is obliged to be associated with entanglement, in contrast to the simple entropy. Thus, the expression (\ref{eq:m1}) implies that the area of the black hole represents how much entanglement is `stored' in the black hole.
In fact, with the quantum state merging protocol \cite{HOW05,HOW07}, we can show that the area of the black hole represents the size of a maximally entangled state which is distillable from an entangled but, in general, {\it mixed} state between positive part $B^+$ inside the black hole and its outside $\bar{B}$ via a process performed by an outside observer $\bar{B}$ without changing the area of the black hole (see Appendix~\ref{sec:A2}).

\section{Discussion}

As shown in this paper, the paradox  \cite{BPZ13,M09,M19}  among a)~the first law of black hole mechanics, b)~the entropy conservation law of quantum mechanics, c)~Hawking's pair-creation picture and d)~the Bekenstein-Hawking equation can be resolved by modifying statement d). That is, we proposed to replace simple entropy in the Bekenstein-Hawking equation (\ref{eq:Beken}) with coherent information as in Eq.~(\ref{eq:m1}). Although it seems to be a conventional view to give up c)~Hawking's pair-creation picture to keep d)~the Bekenstein-Hawking equation, our resolution itself could be compatible even with this view because our area law~(\ref{eq:m1}) cleanly reduces to the Bekenstein-Hawking equation~(\ref{eq:Beken}) unless Hawking radiation is considered (and thus, it could be consistent even with Strominger and Vafa's notable result  \cite{SV96}).

Since the resolution is obtained by applying our equation (\ref{eq:m1}) to {\it changes} of {\it stationary} black holes, any other argument based on the application beyond this manner is merely a conjecture. For instance, it is uncertain whether our equation (\ref{eq:m1}) can apply to the birth and the end of a black hole, because it cannot be treated as being in a stationary state at those times. Nevertheless, it would still be an illustrative exercise to develop our speculations about the life of a black hole further, in order to contextualise our argument (\ref{eq:m1}) in the information loss paradox and the firewall paradox.

If the birth or the `death' (for e.g. by evaporation) of a black hole were connected via a quasi-static process from a stationary black hole, we could determine an integration constant which appears through integrating our equation (\ref{eq:m1}). A simplest way for this is to follow a custom that the integration constant is assumed to be zero, as Hawking has speculated \cite{H76} for Bekenstein-Hawking equation (\ref{eq:Beken}). In our equation (\ref{eq:m1}), this corresponds to the choice that the coherent information $I(\bar{B} \rangle B^+)$ tends to zero as the area $A_B$ tends to zero. 
Then, at the instant of time when an event horizon of a black hole $B$ is formed, we have $(c^3A_B)/(4G \hbar)=S(B^+)$ \cite{SL04}. 
At this point, it is reasonable to assume that the positive part $B^+$ of the black hole $B$ is in a microcanonical ensemble, similarly to a conventional assumption that the Bekenstein-Hawking entropy $S(B)$ of Eq.~(\ref{eq:Beken}) represents the von Neumann entropy of a microcanonical ensemble \cite{H90,SV96,SL04}, i.e., the coarse-grained entropy \cite{AHMST20}. 
This is possible because Hawking radiation has not yet started at that point and thus the black hole $B$ has not yet had any boson $H^-$ with a negative frequency (i.e., $B^-$ is empty at that point), implying that there is no objection to the use of statistical mechanics for the {\it formation} of a black hole.
Hence, the black hole formation follows the second law of thermodynamics, implying that the initial black hole entropy $S(B^+)=(c^3A_B)/(4G \hbar)$ is not less than the entropy of particles that originally formed the black hole upon gravitational collapse \cite{SL04}.
However, once the black hole becomes a stationary state where Hawking radiation starts according to Hawking's pair-creation picture, 
we cannot use conventional statistical mechanics anymore, and it is thus more reasonable to consider that
the black hole follows our equation (\ref{eq:m1}), rather than Bekenstein-Hawking equation (\ref{eq:Beken}). 

If the black hole is located in an environment where Hawking radiation from the hole is larger than absorption by the hole, the area of the black hole is decreasing until its end with $M_B \approx 0$ and $A_B\approx 0$. In this end of the black hole, our equation (\ref{eq:m1}) implies $S(B^+) \approx S(B^-)$. At the same time, $S(B)=S(B^+B^-)=S(\bar{B})$ is huge\footnote{Notice that this entropy $S(B)$ is unaltered even if we have annihilation between $B^+$ and $B^-$ in the form of a unitary interaction, although we do not know whether the unitarity assumption should hold, even if the size of the remnant is comparable with the Planck scale.} if the black hole $B$ is initially large, implying the existence of a remnant \cite{P92,M09}. The gravity made by this remnant is negligibly small because the original positive mass of the black hole is screened for a static observer at infinity by Hawking's negative-frequency
particles inside the remnant hole.
The entanglement problem \cite{M19} points out that it is difficult \cite{P92,M09} to associate such a remnant with a normal object following statistical mechanics, because it has the huge entropy $S(B)$ despite its extremely small mass and area. However, this concern does not apply to our black hole, because it is modelled to be composed not only of normal objects (such as ones composing $B^+$) but also of negative-frequency bosons $H^-$ with unbounded negative-frequency spectra, refuting the use of statistical mechanics. 
Therefore, our current model sheds new light on a `remnant' model that is now free from the the entanglement problem, as well as the causality problem, and thus also from the information loss paradox \cite{M19}.

Although our equation (\ref{eq:m1}) is proposed to provide a reasonable view on conversions between stationary states of a black hole under the existence of Hawking's pair creation, 
there is a model which focuses more on providing a perhaps more reasonable view on the end of a black hole, i.e., its complete evaporation without any remnant. 
In particular, Page gave a notable model \cite{P93a} where a black hole and the Hawking radiation are deemed to {\it always} compose a randomly chosen {\it pure} state \cite{P93b,HP07} on a Hilbert space with a fixed dimension. The dimension of the radiation subsystem increases while the dimension of the black hole subsystem decreases, through Hawking radiation modelled as tunnelling \cite{PW00,BPZ13,BP11}, rather than the pair creation. Combined with a recent view \cite{PSW06} showing the equal a priori probability postulate of conventional statistical mechanics to be unnecessary, Page's model can explain that {\it soon after} the Hawking radiation starts from a huge black hole, the state of the radiation subsystem is close to a canonical ensemble with an extremely low temperature (although the total system is in a pure state), consistent with a conclusion drawn from conventional statistical mechanics which regards the total system in a microcanonical ensemble. However, in contrast to our speculation, even at this point, the black hole subsystem is close to a pure state, rather than a microcanonical ensemble, given its large dimension. Besides, Page's model predicts the late Hawking radiation emerging after the Page time---at which the dimensions of the black hole subsystem and the radiation system are the same---to be in an almost maximally entangled state with the early radiation. This leads to
an inconsistency \cite{AMPS13,M09}---from the monogamy of entanglement \cite{HHHH09}---with Hawking's pair creation picture in which the late radiation should also be fully entangled with modes behind the horizon. Thus, Page's model is incompatible with Hawking's pair-creation picture, in contrast to our model [although there are many other proposals to resolve the firewall paradox without invoking a remnant model (see, e.g., Ref.~\cite{H16})]. On the other hand, Page's model presents a perhaps more reasonable view on the end of the black hole, rather than a remnant model: all the information held by a black hole comes back to our universe after the black hole evaporation. Thus, Page's model is different in many aspects from our speculation about life of a black hole given by our equation (\ref{eq:m1}).

Our area law (\ref{eq:m1}), which identifies that the horizon area of a stationary black hole represents entanglement, would put the use of several tools developed in the field of quantum information to understand black hole physics on a more rigorous footing.
For instance, if we could associate the dynamics to have a black hole through gravitational collapse of a star with a quantum communication protocol over a quantum channel network,
by applying recently derived fundamental upper bounds on obtainable entangled bits between arbitrary two regions over the network \cite{AML16,AK17,BA17,P16,R18,ABCEL21},
we could establish a direct relation between the entanglement, quantified by coherent information $I(\bar{B}\rangle B^+)$ in our equation (\ref{eq:m1}), and the geometric quantity, i.e., the horizon area $A_B$. This would unveil the origin of why a black hole follows an area law---which argues that the entanglement of a region with its outside is upper bounded by its area, rather than its volume (see Appendix~\ref{sec:A3} for more precise explanation about this conjecture). 
\\\\

\section*{Acknowledgment}

We sincerely thank R. Jozsa for speculating about the possibility that the negative-frequency particles `falling into' a black hole may have negative entropy, 
M.~Koashi for having informed K.A. of an interesting implication coming from an analogy between two-mode squeezed states and Hawking radiation ten years ago,
and H.-K. Lo for emphasising the observer dependence of quantum field theory in a curved spacetime.
K.A. is especially thankful to R. Jozsa and the University of Cambridge for giving him an opportunity to stay in R. Jozsa's group during his sabbatical. 
We also thank T. Honjo, M. Hotta, K. Inaba, W. J. Munro, Y. Nakata, K. Shimizu and T. Takayanagi for helpful discussion. 
K.A. thanks support, in part, from PRESTO, JST JP-MJPR1861, from CREST, JST JP-MJCR1671, from Moonshot R\&D, JST JPMJMS2061, from
MEXT-JSPS Grant-in-Aid for Transformative Research Areas (A), No. 21H05183 and from R\&D of ICT Priority Technology (JPMI00316).
S.S. is supported by a Royal Commission for the Exhibition of 1851 Research Fellowship.
Part of this work was done when S.S. was a PhD student at DAMTP, University of Cambridge, supported by the SERB (Government of India) and the Cambridge Trust through a Cambridge-India Ramanujan scholarship.
G.K. thanks support, in part, from the JSPS Kakenhi (C) No. 17K05591, (C) No. 20K03779, (C) No. 21K03388 and JST CREST JPMJCR2113.
Part of this work was done when G.K. belonged to NTT Communication Science Laboratories.


%


\let\doi\relax


\appendix

\section{The area theorem and the generalised second law}\label{sec:A1}

It is instructive to interpret Hawking's area theorem \cite{H72}
and Bekenstein's generalised second law \cite{B74} with the Bekenstein-Hawking equation (3) in the quantum mechanical point of
view, i.e., in a pure quantum information theoretic manner, by assuming that $S$ represents the von Neumann entropy. 
Hawking's area theorem shows that the horizon area $A_B$ satisfies 
\begin{equation}
{\rm d} A_B \ge 0 \label{eq:area}
\end{equation}
for any classical process except for Hawking radiation. Bekenstein's generalised second law argues
that the second law of thermodynamics holds for the sum of the black hole entropy $S(B)$ and 
the entropy $S(\bar{B})$ of matter $\bar{B}$ outside the black hole:
\begin{equation}
{\rm d} S(B)+ {\rm d} S(\bar{B}) \ge 0. \label{eq:gsl}
\end{equation}

The area theorem (\ref{eq:area}) for any `classical' process can be explained with the Bekenstein-Hawking equation (3). For instance, when a particle $C$ with entropy $S(C)$ falls into the black hole, it increases the entropy of the black hole (${\rm d}S(B) \ge 0$), which leads to a corresponding increase in area (${\rm d} A_B \ge 0$) given by Eq.~(\ref{eq:area}), from the Bekenstein-Hawking equation (3).
The increase of area $A_B$ during a black hole merger---which is also expected from the area theorem (\ref{eq:area})---can be understood in this picture in the following manner. Suppose that two black holes $B_1$ and $B_2$ merge together to form a new black hole $B$ through an isometry $\hat{U}_{B_1B_2  \to BG}$---where note that the isometry is merely the application of a unitary operator to system $B_1B_2 R$ with an auxiliary system $R$ in a pure state, according to the standard formalism of quantum information theory. Assume that $B_1$ and $B_2$ are initially decoupled, as expressed by the statement $I(B_1:B_2)=0$. Here 
$G$ is a system emitted to our universe as a back reaction of the merging (e.g. gravitational waves). Then, information theory says that
\begin{align}
S(B)=&S(BG)-S(G|B) =S(B_1B_2)-S(G|B) \nonumber \\
= &S(B_1)+S(B_2) - S(G|B), \label{eq:merge}
\end{align}
where $S(G|B):=S(BG)-S(B)$. Therefore, if the system $G$ is in either a pure state or an entangled state so as to have negative conditional entropy ($S(G|B) \le 0$), the net black hole area increases, in accordance with the Bekenstein-Hawking equation (3). In particular, if initial black holes $B_1$ and $B_2$ and the final black hole $B$ are stationary and the integration constant of the Bekenstein-Hawking equation (3) is zero, from Eq.~(\ref{eq:merge}),
we have $A_{B}\ge A_{B_1} +A_{B_2}$. Since gravitational waves can tell us of the existence of black hole mergers \cite{A16}, it may not be unnatural for the emitted system $G$ to be highly entangled with the black hole $B$.

On the other hand, Bekenstein's generalised second law (\ref{eq:gsl}) can be regarded as a consequence of the following equation, which holds for any unitary dynamics $\hat{U}_{B\bar{B} \to B'\bar{B}'}$ that converts the initial system $B\bar{B}$ into a combined system of the black hole $B'$ and its outside $\bar{B}'$:
\begin{align}
S(B')-S(B) +S(\bar{B}')-S(\bar{B}) 
&= S(B'\bar{B}')+I(B':\bar{B}') - S(B\bar{B}) - I(B:\bar{B})  \nonumber \\
&= I(B':\bar{B}')- I(B:\bar{B}),
\end{align}
where $I$ is the mutual information defined by $I(B:\bar{B}):=S(B)+S(\bar{B})-S(B\bar{B})$. Hence, for any unitary process which increases the mutual information between the black hole and its outside (that is, $I(B':\bar{B}')- I(B:\bar{B})\ge 0$), Eq.~(\ref{eq:gsl}) holds. Notice that this equation, related to the subadditivity of the entropy, is also an underlying basis of quantum thermodynamics \cite{P89,Peres}. We note that the generalised second law has also been proven even in a semi-classical regime, taking into account quantum field theory \cite{Wa12}.
Therefore, the Bekenstein-Hawking equation~(3) could be consistent with any of the phenomena following the area theorem (\ref{eq:area}) or the generalised second law (\ref{eq:gsl}), even from a quantum information theoretic viewpoint.

\section{Quantum information theoretic meaning of our alternative equation}\label{sec:A2}

Here, with the quantum state merging protocol \cite{HOW05,HOW07}, we show that the area $c^3 (4G\hbar)^{-1} A_B$ in Eq.~(4) represents the number of maximally entangled states which are distillable between positive subsystem $B^+$ of the black hole and its outside $\bar{B}$ via a process of an outside observer $\bar{B}$ without changing the area. 
In quantum information theory,
the coherent information $I(\bar{B} \rangle B^+)$ has a clear operational meaning in the quantum state merging \cite{HOW05,HOW07}. In particular, for $I(\bar{B} \rangle B^+)\ge 0$, the coherent information $I(\bar{B} \rangle B^+)$ represents the distillable entanglement by merging the state of $\bar{B}$ into $B^+$ with local operations and classical communication. An optimal merging protocol is given as follows: Given $\bar{B}B^+B^-=\bar{B}B_1^+B_0^+B^-$ (i.e., $B^+=B_1^+B_0^+$) initially in a pure state $\ket{\Psi}_{\bar{B}B^+B^-}=:\ket{\Psi}_{\bar{B}B_1^+B^-} \ket{0}_{B_0^+}$ with an ancilla state $\ket{0}_{B_0^+}$ and $\ket{\Psi}_{\bar{B}B^+_{1}B^-}= \bigotimes_{i=1}^n \ket{\psi}_{\bar{b}_i b_i^+b_i^-}$ of $n (\gg 1)$ copies of elementary systems $\bar{b}_1 \bar{b}_2 \cdots \bar{b}_n (= \bar{B})$ and $b_1^\pm b_2^\pm \cdots b_n^\pm (= B^\pm_1)$, we perform a unitary operation $\hat{T}_{\bar{B}}$ on $\bar{B}$ chosen at random according
to the uniform measure (Haar measure), followed by projections $\{\hat{P}_{\bar{B}\to \bar{B}_1}^{(k)}\}_k$ onto a subspace $\bar{B}_1$ of $\bar{B}$ with dimension $\simeq e^{I(\bar{B}\rangle B^+)}$ (or less in general). Then, for almost all the outcomes $k$, this protocol provides a state $\ket{\Psi_k}_{\bar{B}_1 B_1^+ B^-}\ket{0}_{B_0^+}$ close to $\hat{V}^{(k)}_{B^+} \ket{\Phi^+}_{\bar{B}_1B_2^+}  \ket{\Psi}_{B_3^+ B_1^+B^-}$ with probability $p_k$, where 
$\hat{V}^{(k)}_{ B^+}$ is a unitary operator on $B^+=B_1^+ B_0^+ =B_1^+ B_2^+ B_3^+ $, $\ket{\Phi^+}_{\bar{B}_1B_{2}^+} $ is a maximally entangled state and $\ket{\Psi}_{B_3^+ B_1^+ B^-} $ is the merged state defined as $\ket{\Psi}_{B_3^+ B_1^+ B^-} :=\hat{1}_{\bar{B} \to B_{3}^+} \ket{\Psi}_{\bar{B}B_{1}^+B^-}$. This means that we can consider a coherent version of the merging protocol, by using an isometry $\hat{U}_{\bar{B}\to \bar{B}_1\bar{B}_2 }:=\sum_k \hat{P}_{\bar{B}\to \bar{B}_1}^{(k)} \ket{k}_{\bar{B}_2}\hat{T}_{\bar{B}}$ with a subsystem $\bar{B}_2$ of $\bar{B}$ different from $\bar{B}_1$ and phases $\phi_k$ such that
\begin{align}
\hat{U}_{\bar{B}\to \bar{B}_1\bar{B}_0 } \ket{\Psi}_{\bar{B}B_{1}^+B^-} \ket{0}_{B_0^+} 
=&  \sum_{k}  \sqrt{p_k}\ket{k}_{\bar{B}_2} \ket{\Psi_k}_{\bar{B}_1 B_1^+ B^-}\ket{0}_{B_2^+}
\nonumber \\
\simeq&  \sum_{k} e^{i \phi_k} \sqrt{p_k}\ket{k}_{\bar{B}_2} \hat{V}^{(k)}_{ B^+}  \ket{\Phi^+}_{\bar{B}_1B_{2}^+}  \ket{\Psi}_{B_{3}^+ B^+_1 B^-}   \nonumber  \\
= & \hat{W}_{\bar{B}_2 B^+ } \ket{\varphi}_{\bar{B}_2} \ket{\Phi^+}_{\bar{B}_1B_{2}^+}  \ket{\Psi}_{B_{3}^+ B^+_1 B^-},
\label{eq:mer}
\end{align}
where $\{\ket{k}_{\bar{B}_2}\}$ is a set of orthonormal states, $\ket{\varphi}_{\bar{B}_2} = \sum_{k} e^{i \phi_k} \sqrt{p_k}\ket{k}_{\bar{B}_2}$, and $\hat{W}_{\bar{B}_2 B^+ }$ is a controlled unitary operator defined by $\hat{W}_{\bar{B}_2 B^+}  :=\sum_k \ket{k}\bra{k}_{\bar{B}_2} \otimes \hat{V}^{(k)}_{ B^+}  $. 

In our context, a coherent version of the state merging protocol suggests the following:  in principle, an observer $\bar{B}$ outside the black hole $B$ can convert the initial state $\ket{\Psi}_{\bar{B}B^+B^-}$ into a standard state which is of almost the same area but maximally entangled only with the positive part $B^+$ of the black hole. In fact, as shown by Eq.~(\ref{eq:mer}), if the observer performs the isometry $\hat{U}_{\bar{B}\to \bar{B}_1\bar{B}_2 }$ on the initial state $\ket{\Psi}_{\bar{B}B^+B^-}$ and then throws the system $\bar{B}_2$ into the black hole as its new member $B_4^+$, the total system is approximately in the standard state $  \ket{\varphi}_{B_4^+}\ket{\Phi^+}_{\bar{B}_1B_{2}^+} \ket{\Psi}_{B_{3}^+ B^+_1  B^-} $ up to the freedom of local unitaries (corresponding to $ \hat{W}_{\bar{B}_2 B^+}$ in Eq.~(\ref{eq:mer})) on positive subsystem $B^+_1 B^+_2  B_{3}^+B_{4}^+ $ of the black hole. 
Hence, the area of the black hole $ B^+_1 B_2^+ B_{3}^+B_{4}^+B^-$ is $I(\bar{B}_1 \rangle B^+_1 B_2^+ B_{3}^+B_{4}^+)  = S(B_2^+) \simeq I(\bar{B}\rangle B^+)$. That is,
although this coherent merging process includes throwing of a system into the black hole, its area does not change, representing the dimension of the maximally entangled state $\ket{\Phi^+}_{\bar{B}_1B_{2}^+} $ that is distillable between the positive subsystem $B^+$ of the black hole and its outside.

\section{Conjecture on a possible information theoretic reason on the area law}\label{sec:A3}

We conjecture a possible information theoretic reason on why a black hole follows an area law like Eq.~(4), which argues that the entanglement of a region with its outside is upper bounded by its area, rather than its volume. Since any motion of any physical system in the spacetime happens along a quantum channel, the gravitational collapse of a star to form a black hole $B$ should be associated with the transmission of physical systems through quantum channels $\{{\cal N}_e\}_{e}$ in a quantum network spread over the spacetime. Then, if we quantify the entanglement stored in a black hole $B$ with the coherent information as in Eq.~(4), this quantity is upper bounded as
\begin{align}
I(\bar{B}\rangle B^+) \le E_{\rm sq} (\bar{B}: B^+) \le  \sum_{e \in {\partial B}} \bar{l}_e E_{\rm sq} ({\cal N}_e),
\end{align}
irrespective of any detail of its dynamics \cite{AML16}. Here $E_{\rm sq}(\bar{B}: B^+)$ represents the squashed entanglement \cite{CW04}, $E_{\rm sq} ({\cal N}_e)$ is the squashed entanglement \cite{TGW14} of the channel ${\cal N}_e$, $\bar{l}_e$ represents how many times (on average) the channel ${\cal N}_e$ has been used in the process to form the entanglement \cite{AK17}, and the summation is taken over all $e \in \partial B$ which specify channels ${\cal N}_{e}$ connecting the inside and the outside of the black hole $B$ being formed, across its horizon. If the gravitational collapse satisfies $\bar{l}_e \le C$ for a constant $C$ (which is related with the speed of the gravitational collapse), this inequality reduces to
\begin{align}
I(\bar{B}\rangle B^+) \le C \sum_{e \in {\partial B}} E_{\rm sq} ({\cal N}_e)= C E_{\rm sq}\left(\bigotimes_{e\in \partial B} {\cal N}_e\right)
\end{align}
by the additivity of the squashed entanglement. Since $E_{\rm sq}\left(\bigotimes_{e\in \partial B} {\cal N}_e\right)  $ is related with the capacity of the channel $\bigotimes_{e\in \partial B} {\cal N}_e$ connecting the inside and the outside of the event horizon in the spacetime, it could be upper bounded by the geometric area \a{$A_B$} of the black hole $B$. This way, we could identify the origin of why a black hole follows the area law as the speed of the gravitational collapse.

\end{document}